# Two-level system loss characterization of NbTi superconducting resonators on Si/SiO$_2$ substrates

Bongkeon Kim[1] and Yong-Joo Doh[1*]

[1]Department of Physics and Photon Science, Gwangju Institute of Science and Technology, Gwangju, 61005, Korea.

## Abstract

Superconducting coplanar waveguide (SCPW) resonators, key components for quantum computing and sensing applications, require a high internal quality factor ($Q_i$) for effective qubit readout and quantum sensing applications. Minimizing two-level system (TLS) losses, particularly at material interfaces, is critical for gatemon and topological qubits operating at low temperatures and in high magnetic fields. NbTi, a superconducting alloy with a high upper critical field, enables SCPW resonators resilient to such conditions. We fabricated NbTi SCPW resonators on Si/SiO$_2$ substrates and systematically characterized their TLS-limited quality factors as functions of temperature and microwave photon number. Our results demonstrate that NbTi-based SCPWs on Si/SiO$_2$ substrates provide a promising platform for developing next-generation quantum circuits.

**Corresponding Author**

*E-mail: yjdoh@gist.ac.kr

## 1. Introduction

Superconducting coplanar waveguide (SCPW) resonators consist of transmission lines with a central conductor and two ground planes [1]. These resonators, when coupled to superconducting qubits, enable quantum state readout [2, 3] and quantum sensor applications [4, 5]. Their performance is characterized by the internal quality factor ($Q_i$), defined as the inverse of microwave loss [6]. Since high $Q_i$ means low microwave loss, the development of high-$Q_i$ resonators is essential for advancing quantum computing and sensing technologies. In particular, defects in amorphous dielectrics or at the substrate–superconductor interface interact with microwave photons, resulting in two-level system (TLS) loss [7, 8]. In the low-temperature and single-photon regime relevant to superconducting qubits, TLS loss dominates over other loss mechanisms [8].

Gatemon qubits, in which the qubit's properties are controlled using electrostatic gate voltages rather than magnetic flux, enable electrical tuning of both qubit frequency and coupling, offering notable advantages in tunability and scalability [9, 10]. To prevent gate leakage current, gatemon qubits are typically fabricated on a silicon substrate covered with a silicon dioxide layer [9, 11, 12]. However, $SiO_2$, as an amorphous insulator, is known to exhibit significant TLS loss that can degrade qubit performance [8, 9]. Moreover, a gate-tunable Majorana-transmon qubit [13, 14], which is a gatemon qubit incorporating topologically protected Majorana zero modes [15] requires an external magnetic field of approximately 1 T for its operation [16]. Thus a SCPW resonator with robust magnetic field resilience is essential for realizing gate-tunable topological qubits [17].

NbTi is a promising superconducting alloy for fabricating magnetic-field-resilient resonators, with an upper critical magnetic field exceeding 9 T [18]. In this study, we fabricated NbTi SCPW resonators on Si/$SiO_2$ substrates and measured $Q_i$ and resonance frequency ($f_c$) as functions of temperature and input microwave power. The extracted TLS-limited quality factor ($Q_{TLS,0}$) exceeded those of Nb- or Al-based superconducting resonators [19-22]. These results demonstrate that NbTi SCPW resonators provide a promising platform for realizing gatemon and topological qubits.

## 2. Experimental methods

The NbTi SCPW resonators were fabricated on undoped silicon (100) wafers with a room-temperature resistivity of $\rho_{Si} = 10$ kΩ·cm, coated with a 150-nm-thick thermally grown SiO$_2$ layer. NbTi films were deposited using a pulsed DC sputtering system with a high-purity (99.9%) NbTi alloy target (50/50 at.%, iTASCO). During deposition, the Ar pressure was maintained at 2.0 mTorr and the sputtering power at 200 W, yielding a deposition rate of 30 nm/min. The resonators were patterned using UV photolithography and etched by reactive ion etching in an SF$_6$/O$_2$ atmosphere with flow rates of 4 sccm/1 sccm. Residual photoresist was removed using mr-Rem 700 (Micro Resist Technology GmbH) to complete the resonator fabrication [23]. Optical microscopy images of the SCPW resonators are shown in Figs. 1a-d.

Each device contains either six (D1) or two (D2) quarter-wavelength (λ/4) SCPW resonators. These resonators, highlighted in purple in Fig. 1a, have different lengths ($l$) and are inductively coupled to a feedline (shown in green in Fig. 1b) through a hanger-type geometry. Each resonator has one end shorted to the ground plane at the coupling point (Fig. 1c) and the opposite end left open. The key design parameters are summarized in Table 1, including central conductor width ($s$), slot spacing ($w$), geometrical capacitance ($C_g$), geometrical inductance ($L_g$), kinetic inductance ($L_k$), and characteristic impedance ($Z_0$). The ground plane surrounding the resonators is patterned with hole arrays that trap Abrikosov vortices induced by external magnetic fields. Device D1 uses $5.8 \times 5.8$ μm$^2$ square holes with 5.8 μm spacing, while device D2 uses $4.0 \times 4.0$ μm$^2$ with 4.0 μm spacing. The resonant frequencies $f_c$ range from 3.95 to 4.12 GHz for D1 and from 4.95 to 5.38 GHz for D2. Using these $f_c$ values, the kinetic inductance $L_k = 1/(16l^2f_c^2C_g) - L_g$ and the kinetic inductance ratio $\alpha_k = L_k/(L_g+L_k)$ are calculated in Table 1.

After device fabrication, the resonator sample was bonded to a PCB sample holder using a wedge bonder, as shown in Fig. 2a. The assembled holder was subsequently enclosed in an aluminum box to minimize losses from quasiparticles and magnetic fields. The aluminum box was then mounted onto the mixing-chamber plate of a dilution refrigerator (LD400, Bluefors Oy) and cooled to a base temperature of 8 mK (see Fig. 2b). The insertion loss ($|S_{21}|$) of the sample was measured using a vector network analyzer (VNA; PNA-L N5231A, Keysight Technologies Inc.) (see Fig. 2c). The input microwave signal was attenuated by a total of 60 dB within the refrigerator, while the output signal was routed through two cryogenic isolators (QCI-G0401201AU, Quinstar Technology Inc.) connected in series. The transmitted microwave signal was then amplified using a high electron mobility transistor (HEMT) amplifier (LNF-LNC0.3_14B, Low Noise Factory AB) at 4 K and a low-noise amplifier (LNA; L3 AFS3-00101200-35-ULN-R, Narda-MITEQ) at room temperature.

### 3. Results and discussion

The $|S_{21}|$ characteristic curves measured from devices D1 and D2 are shown in Figs. 3a and 3b, respectively. The applied microwave power was $P_{in}$ = -115 dBm. Each resonance dip occurs at the resonant frequency $f_c$. The transmission coefficient ($|S'_{21}|$) curves for resonators D1R3 and D2R1 are shown in Figs. 3c and 3d, respectively, which were obtained from $|S_{21}|$ using the relation of $|S'_{21}| = 10^{(|S_{21}| - LL)/20}$ with correction for the coaxial-line loss (LL). The complex transmission coefficient $S'_{21}$ is expressed by the following equation [24]:

$$S'_{21} = A \left(1 + \alpha' \frac{f - f_c}{f_c}\right) \left(1 - \frac{Q_l e^{i\theta} / |Q_e|}{1 + 2i Q_l (f - f_c)/f_c}\right) \qquad (1).$$

Here, $A$ denotes the amplitude parameter, $\alpha'$ means the background slope, $Q_e = |Q_e|e^{i\theta}$ is the complex external quality factor, which is related to the coupling quality factor $Q_c$ through $1/Q_c$ = Re($1/Q_e$). The loaded quality factor is given by $1/Q_l = 1/Q_c + 1/Q_i$. The solid curves in Figs. 3c and 3d represent the theoretical fits using Eq. (1), resulting $Q_i = 4.95 \times 10^4$ and $Q_c = 4.37 \times 10^3$ for resonator D1R3, and $Q_i = 1.07 \times 10^5$ and $Q_c = 9.52 \times 10^3$ for resonator D2R1. Additional fitting parameters are summarized in Table 1. We note that the average photon numbers ($n_{ph}$) within the resonators are $n_{ph} = 3.5 \times 10^2$ for D1R3 and $5.2 \times 10^2$ for D2R1, calculated using the equation of $n_{ph} = \frac{2}{h f_c^2} \frac{Q_l^2}{Q_c} P_{in}$ [25], where $h$ is Planck's constant.

The $|S_{21}|$ measurements as a function of photon number are displayed in Figs. 4a and 4b for resonators D1R3 and D2R1, respectively. Both resonators maintained a stable resonance frequency over six orders of magnitude in $n_{ph}$, while the resonance dip exhibited increasing depth with higher $n_{ph}$. Similar behavior was observed in other devices. Quantitative fitting of the data using Eq. (1) indicates that the internal quality factor increases with $n_{ph}$ and approaches saturation at large photon numbers, as shown in Figs. 4c-f. This behavior is attributed to the saturation of two-level systems (TLSs) [19, 26]. At low photon numbers, TLSs residing in the substrate and interfacial regions of the resonator absorb photons, thereby reducing $Q_i$. With increasing $n_{ph}$, the stronger microwave field drives TLSs into saturation, suppressing their contribution to energy dissipation and enhancing $Q_i$ [26]. At sufficiently large $n_{ph}$, however, other loss channels become dominant, leading to the saturation of $Q_i$. A detailed quantitative analysis is presented below.

In the TLS model, the total loss in SCPW resonators can be decomposed into TLS-induced loss ($1/Q_{TLS}$) and other residual loss channels ($1/Q_{other}$) [26]:

$$\frac{1}{Q_i(n_{ph},T)} = \frac{1}{Q_{TLS,0}} \frac{\tanh\ (hf_c/2k_BT)}{\sqrt{1+(n_{ph}/n_s)^\beta}} + \frac{1}{Q_{other}} \qquad (2),$$

where $Q_{TLS,0}$ denotes the intrinsic TLS-limited quality factor at zero photon number, $k_B$ is the Boltzmann's constant, $n_s$ is the characteristic photon number associated with TLS saturation, and $\beta$ is a fit parameter. Figures 4c-f show that the fitted curves (red lines) based on Eq. (2) agrees well with the experimental data (solid circles). The extracted fit parameters are summarized in Table 1. From the fits, we obtain $Q_{TLS,0} = 7.61 \times 10^4$ for the D1R3 resonator and $9.21 \times 10^4$ for the D2R1 resonator, corresponding to TLS-related loss tangents of $\delta = 1/Q_{TLS,0} = 1.31 \times 10^{-5}$ and $1.09 \times 10^{-5}$, respectively.

The temperature dependence of $Q_i$ measured at various photon numbers is shown in Figs. 5a-d for both D1R3 and D2R1 resonators. For both devices, $Q_i$ remains nearly constant below 0.1 K, but increases notably as temperature rises from 0.1 K to 1 K. This behavior occurs because thermal fluctuations depolarize the electric dipoles of TLS, thereby reducing photon loss caused by TLS defects [25, 26]. Moreover, $Q_i$ exhibits pronounced temperature dependence at low $n_{ph}$, in contrast to the minimal temperature dependence observed at high $n_{ph}$. At low $n_{ph}$, TLS defects remain unsaturated and thermal fluctuations become the dominant source of resonator loss [26]. However, once the microwave field sufficiently saturates TLS defects at high $n_{ph}$, thermal effects no longer significantly affect the loss mechanism. Fitting the data in Figs. 5a and 5b using Eq. (2) yields $Q_{TLS,0} = (6.61 \pm 2.04) \times 10^4$ for D1R3 and $(5.98 \pm 1.14) \times 10^4$ for D2R1, averaged over different photon numbers. These values are comparable to those obtained from $Q_i$ measured at 8 mK as a function of $n_{ph}$.

As temperature increases above 1 K, the saturated $Q_i$ values decrease monotonically (Figs. 5c and 5d). This reduction in $Q_i$ is attributed to the increase in thermal quasiparticle density at higher temperatures, as predicted by the Mattis-Bardeen (MB) theory [27-29]. According to MB theory, temperature-dependence of $Q_i$ is described by:

$$\frac{1}{Q_{i,qp}(T)} = \alpha_{k,Q}(T)\frac{\sigma_1(T) - \sigma_1(0)}{\sigma_2(T)} \qquad (3).$$

Here, $\alpha_{k,Q}$ is the kinetic inductance fraction, and $\sigma_1(T)$ and $\sigma_2(T)$ are the real and imaginary parts of the complex conductivity $\sigma = \sigma_1 - i\sigma_2$ of the superconducting film, respectively, which are given by:

$$\frac{\sigma_1}{\sigma_n} = \frac{4\Delta(T)}{hf_c}e^{-\frac{\Delta(T)}{k_BT}}\sinh\left(\frac{hf_c}{2k_BT}\right)K_0\left(\frac{hf_c}{2k_BT}\right) \qquad (4)$$

$$\frac{\sigma_2}{\sigma_n} = \frac{\pi\Delta(T)}{hf_c}\left[1 - e^{-\frac{\Delta(T)}{k_BT}}e^{-\frac{hf_c}{2k_BT}}I_0\left(\frac{hf_c}{2k_BT}\right)\right] \qquad (5),$$

where $\sigma_n$ is the normal-state conductivity, $\Delta(T)$ is the temperature-dependent superconducting energy gap, and $I_0(x)$ and $K_0(x)$ are the modified Bessel functions of the first and second kind, respectively. The MB theory fits show good agreement with the experimental data (see Figs. 5c and 5d). From the theoretical calculations, the kinetic inductance fractions were obtained as $\alpha_{k,Q} = 0.182$ for D1R3 and 0.201 for D2R1, respectively. These values are consistent with $\alpha_k$ determined from $f_c$ at 8 mK, as listed in Table 1.

The TLS-limited quality factor can also be inferred from the resonance frequency. Figures 6a and 6b present the temperature dependence of $f_c$, where the fractional resonance frequency, $\delta f_c(T)/f_c = [f_c(T) - f_c(8\text{ mK})]/f_c(8\text{ mK})$, is plotted. Unlike $Q_i$, $f_c$ is almost insensitive to $n_{ph}$; temperature is the primary factor to influence its behavior. Below 1 K, $f_c(T)$ decreases as temperature decreases, which is attributed to changes in the effective permittivity induced by TLS defects [30]. As temperature decreases, TLS defects relax into their ground state and become increasingly polarized, thereby raising the effective permittivity. This increase in permittivity leads to higher capacitance, resulting in a red shift of $f_c$.

The temperature dependence of $f_c$ arising from TLS defects can be described by [30]:

$$\frac{\delta f_c(T)}{f_c} = \frac{1}{\pi Q_{TLS,0}}Re\left[\Psi\left(\frac{1}{2} + i\frac{hf_c(8\ m\ K)}{2\pi k_BT}\right) - \ln\left(\frac{hf_c(8\ m\ K)}{2\pi k_BT}\right)\right] \qquad (6)$$

where $\Psi(x)$ is the complex digamma function. The fit results using Eq. (6), shown as solid lines in Figs. 6a and 6b, exhibit excellent agreement with the experimental data. The TLS-limited quality factors derived from $f_c$ are $Q_{TLS,0} = 6.40 \times 10^4$ for D1R3 and $8.02 \times 10^4$ for D2R1, which are consistent with those determined from the $n_{ph}$ and temperature dependence of $Q_i$. Averaging the TLS quality factors obtained from three different methods yields $Q_{TLS,0,avg} = (6.87 \pm 0.65) \times 10^4$ for D1R3 and $(7.74 \pm 1.63) \times 10^4$ for D2R1, corresponding to loss tangents of $\delta_{avg} = 1.46 \times 10^{-5}$ and $1.29 \times 10^{-5}$, respectively. It is noteworthy that our averaged $Q_{TLS,0}$ values exceed those previously reported for Si/SiO$_2$ substrates by more than an order of magnitude [19], and represent the highest values achieved for Si/SiO$_2$ substrates to date [20-22]. In addition, these results predict a TLS-limited relaxation time of $T_1 = Q_{TLS,0}/2\pi f = 2.9$ μs for qubits operating at 5 GHz, which is much longer than the 1.3 μs reported for single gatemon qubits on Si/SiO$_2$ [12]. This demonstrates that NbTi resonators on Si/SiO$_2$ substrates are promising platforms for gatemon and Majorana-transmon qubit applications.

Above 1 K, $f_c$ decreases markedly as temperature increases (see Figs. 6c and 6d). This behavior is attributed to the generation of thermal quasiparticles [27, 29]. As temperature increases, Cooper pairs in the superconducting film break, causing the London penetration depth ($\lambda_L$) to increase. Since the kinetic inductance in thin films is proportional to $\lambda_L^2$ [31], a larger $\lambda_L$ results in a higher $L_k$ and thus a reduction in $f_c$. This red shift is described by MB theory as [27-29]:

$$\frac{\delta f_c(T)}{f_c} = \frac{\alpha_{k,f}(T)}{2} \frac{\sigma_2(T) - \sigma_2(0)}{\sigma_2(T)} \qquad (7)$$

The fits to Eq. (7), shown as solid lines in Figs. 6c and 6d, agree well with the experimental results. The kinetic inductance fractions used in the theoretical calculations are $\alpha_{k,f} = 0.174$ for D1R3 and 0.168 for D2R1, respectively. These values are consistent with those obtained from the temperature dependence of $Q_i$ and with $\alpha_k$ determined from $f_c$. Therefore, the averaged fractions are $\alpha_{k,avg} = 0.185 \pm 0.013$ for D1R3 and $0.194 \pm 0.023$ for D2R1, closely matching reported values for MoRe resonators of similar thickness [29].

## 4. Conclusion

In this study, we fabricated NbTi SCPW resonators on Si/SiO$_2$ substrates and investigated their loss characteristics at low temperatures. By analyzing the temperature dependence of $Q_i$ and $f_c$ across various photon number regimes, we extracted TLS-limited quality factors ranging from $Q_{TLS,0} = 5.14 \times 10^4$ to $9.21 \times 10^4$. To the best of our knowledge, these values represent the highest TLS-limited quality factors reported to date for superconducting resonators on Si/SiO$_2$ substrates. Our results demonstrate that NbTi resonators on Si/SiO$_2$ substrates would provide a promising platform for developing advanced quantum devices, including gatemon and topological qubits.

## Author contributions

**Bongkeon Kim:** Conceptualization, Device fabrications, Low-temperature measurements, Data analysis and calculation, Visualization, Writing. **Yong-Joo Doh:** Conceptualization, Project administration, Supervision, Validation, Writing.

## Declaration of competing interest

The authors declare that they have no known competing financial interests or personal relationships that could have appeared to influence the work reported in this paper.

## Acknowledgement

This study was supported by the NRF of Korea through the Basic Science Research Program (RS-2018-NR030955, RS-2023-00207732, RS-2025-02317602), the ITRC program (IITP-2025-RS-2022-00164799) funded by the Ministry of Science and ICT, and the GIST Research Project grant funded by the GIST in 2025.

## Data availability

All data in this study are available on request.

# References

[1] C.P. Wen, Coplanar Waveguide: A Surface Strip Transmission Line Suitable for Nonreciprocal Gyromagnetic Device Applications, IEEE Trans. Microw. Theor. Tech., 17 (1969) 1087-1090.

[2] A. Blais, R.-S. Huang, A. Wallraff, S.M. Girvin, R.J. Schoelkopf, Cavity quantum electrodynamics for superconducting electrical circuits: An architecture for quantum computation, Phys. Rev. A, 69 (2004) 062320.

[3] A. Wallraff, D.I. Schuster, A. Blais, L. Frunzio, R.S. Huang, J. Majer, S. Kumar, S.M. Girvin, R.J. Schoelkopf, Strong coupling of a single photon to a superconducting qubit using circuit quantum electrodynamics, Nature, 431 (2004) 162-167.

[4] M. Bal, C. Deng, J.L. Orgiazzi, F.R. Ong, A. Lupascu, Ultrasensitive magnetic field detection using a single artificial atom, Nat. Commun., 3 (2012) 1324.

[5] Y. Kubo, C. Grezes, A. Dewes, T. Umeda, J. Isoya, H. Sumiya, N. Morishita, H. Abe, S. Onoda, T. Ohshima, V. Jacques, A. Dréau, J.F. Roch, I. Diniz, A. Auffeves, D. Vion, D. Esteve, P. Bertet, Hybrid Quantum Circuit with a Superconducting Qubit Coupled to a Spin Ensemble, Phys. Rev. Lett., 107 (2011) 220501.

[6] D.M. Pozar, Microwave engineering: theory and techniques, 2nd ed., John wiley & sons, 1998.

[7] S. Hunklinger, W. Arnold, S. Stein, R. Nava, K. Dransfeld, Saturation of the ultrasonic absorption in vitreous silica at low temperatures, Phys. Lett. A, 42 (1972) 253-255.

[8] J.M. Martinis, K.B. Cooper, R. McDermott, M. Steffen, M. Ansmann, K.D. Osborn, K. Cicak, S. Oh, D.P. Pappas, R.W. Simmonds, C.C. Yu, Decoherence in Josephson Qubits from Dielectric Loss, Phys. Rev. Lett., 95 (2005) 210503.

[9] T.W. Larsen, K.D. Petersson, F. Kuemmeth, T.S. Jespersen, P. Krogstrup, J. Nygård, C.M. Marcus, Semiconductor-Nanowire-Based Superconducting Qubit, Phys. Rev. Lett., 115 (2015) 127001.

[10] G. de Lange, B. van Heck, A. Bruno, D.J. van Woerkom, A. Geresdi, S.R. Plissard, E.P.A.M. Bakkers, A.R. Akhmerov, L. DiCarlo, Realization of Microwave Quantum Circuits Using Hybrid Superconducting-Semiconducting Nanowire Josephson Elements, Phys. Rev. Lett., 115 (2015) 127002.

[11] M. Mergenthaler, A. Nersisyan, A. Patterson, M. Esposito, A. Baumgartner, C. Schönenberger, G.A.D. Briggs, E.A. Laird, P.J. Leek, Circuit Quantum Electrodynamics with Carbon-Nanotube-Based Superconducting Quantum Circuits, Phys. Rev. Appl., 15 (2021) 064050.

[12] H. Zheng, L.Y. Cheung, N. Sangwan, A. Kononov, R. Haller, J. Ridderbos, C. Ciaccia, J.H. Ungerer, A. Li, E.P.A.M. Bakkers, A. Baumgartner, C. Schönenberger, Coherent Control of a Few-Channel Hole Type Gatemon Qubit, Nano Lett., 24 (2024) 7173-7179.

[13] E. Ginossar, E. Grosfeld, Microwave transitions as a signature of coherent parity mixing effects in the Majorana-transmon qubit, Nat. Commun., 5 (2014) 4772.

[14] T.B. Smith, M.C. Cassidy, D.J. Reilly, S.D. Bartlett, A.L. Grimsmo, Dispersive Readout of Majorana Qubits, PRX Quantum, 1 (2020) 020313.

[15] A.Y. Kitaev, Unpaired Majorana fermions in quantum wires, Phys.-Usp., 44 (2001) 131.

[16] V. Mourik, K. Zuo, S.M. Frolov, S.R. Plissard, E.P.A.M. Bakkers, L.P. Kouwenhoven, Signatures of Majorana Fermions in Hybrid Superconductor-Semiconductor Nanowire Devices, Science, 336 (2012) 1003-1007.

[17] M. Aghaee, A. Alcaraz Ramirez, Z. Alam, R. Ali, M. Andrzejczuk, A. Antipov, M. Astafev, A. Barzegar, B. Bauer, J. Becker, U.K. Bhaskar, A. Bocharov, S. Boddapati, D. Bohn, J. Bommer, L. Bourdet, A. Bousquet, S. Boutin, L. Casparis, B.J. Chapman, S. Chatoor, A.W. Christensen, C. Chua, P. Codd, W. Cole, P. Cooper, F. Corsetti, A. Cui, P. Dalpasso, J.P. Dehollain, G. de Lange, M. de Moor, A. Ekefjärd, T. El Dandachi, J.C. Estrada Saldaña, S. Fallahi, L. Galletti, G. Gardner, D. Govender, F. Griggio, R. Grigoryan, S. Grijalva, S. Gronin, J. Gukelberger, M. Hamdast, F. Hamze, E.B. Hansen, S. Heedt, Z. Heidarnia, J. Herranz Zamorano, S. Ho, L. Holgaard, J. Hornibrook, J. Indrapiromkul, H. Ingerslev, L. Ivancevic, T. Jensen, J. Jhoja, J. Jones, K.V. Kalashnikov, R. Kallaher, R. Kalra, F. Karimi, T. Karzig, E. King, M.E. Kloster, C. Knapp, D. Kocon, J.V. Koski, P. Kostamo, M. Kumar, T. Laeven, T. Larsen, J. Lee, K. Lee, G. Leum, K. Li, T. Lindemann, M. Looij, J. Love, M. Lucas, R. Lutchyn, M.H. Madsen, N. Madulid, A. Malmros, M. Manfra, D. Mantri, S.B. Markussen, E. Martinez, M. Mattila, R. McNeil, A.B. Mei, R.V. Mishmash, G. Mohandas, C. Mollgaard, T. Morgan, G. Moussa, C. Nayak, J.H. Nielsen, J.M. Nielsen, W.H.P. Nielsen, B. Nijholt, M. Nystrom, E. O'Farrell, T. Ohki, K. Otani, B. Paquelet Wütz, S. Pauka, K. Petersson, L. Petit, D. Pikulin, G. Prawiroatmodjo, F. Preiss, E. Puchol Morejon, M. Rajpalke, C. Ranta, K. Rasmussen, D. Razmadze, O. Reentila, DJ. Reilly, Y. Ren, K. Reneris, R. Rouse, I. Sadovskyy, L. Sainiemi, I. Sanlorenzo, E. Schmidgall, C. Sfiligoj, M.B. Shah, K. Simoes, S. Singh, S. Sinha, T. Soerensen, P. Sohr, T. Stankevic, L. Stek, E. Stuppard, H. Suominen, J. Suter, S. Teicher, N. Thiyagarajah, R. Tholapi, M. Thomas, E. Toomey, J. Tracy, M. Turley, S. Upadhyay, I. Urban, K. Van Hoogdalem, D.J. Van Woerkom, D.V. Viazmitinov, D. Vogel, J. Watson, A. Webster, J. Weston, G.W. Winkler, D. Xu, C.K. Yang, E. Yucelen, R. Zeisel, G. Zheng, J. Zilke, Q. Microsoft Azure, Interferometric single-shot parity measurement in InAs–Al hybrid devices, Nature, 638 (2025) 651-655.

[18] L. Bottura, A practical fit for the critical surface of NbTi, IEEE Trans. Appl. Supercond., 10 (2000) 1054-1057.

[19] A.D. O'Connell, M. Ansmann, R.C. Bialczak, M. Hofheinz, N. Katz, E. Lucero, C. McKenney, M. Neeley, H. Wang, E.M. Weig, A.N. Cleland, J.M. Martinis, Microwave dielectric loss at single photon energies and millikelvin temperatures, Appl. Phys. Lett., 92 (2008) 112903.

[20] J.M. Sage, V. Bolkhovsky, W.D. Oliver, B. Turek, P.B. Welander, Study of loss in superconducting coplanar waveguide resonators, J. Appl. Phys., 109 (2011) 063915

[21] T. Lindström, J.E. Healey, M.S. Colclough, C.M. Muirhead, A.Y. Tzalenchuk, Properties of superconducting planar resonators at millikelvin temperatures, Phys. Rev. B, 80 (2009) 132501.

[22] C.X. Yu, S. Zihlmann, G. Troncoso Fernández-Bada, J.-L. Thomassin, F. Gustavo, É. Dumur, R. Maurand, Magnetic field resilient high kinetic inductance superconducting niobium nitride coplanar waveguide resonators, Appl. Phys. Lett., 118 (2021) 054001.

[23] B. Kim, M. Jung, J. Kim, J. Suh, Y.-J. Doh, Fabrication and characterization of superconducting

coplanar waveguide resonators, Prog. Supercond. Cryogenics, 22 (2020) 10-13.

[24] M.S. Khalil, M.J.A. Stoutimore, F.C. Wellstood, K.D. Osborn, An analysis method for asymmetric resonator transmission applied to superconducting devices, J. Appl. Phys., 111 (2012) 054510.

[25] A. Bruno, G. de Lange, S. Asaad, K.L. van der Enden, N.K. Langford, L. DiCarlo, Reducing intrinsic loss in superconducting resonators by surface treatment and deep etching of silicon substrates, Appl. Phys. Lett., 106 (2015) 182601.

[26] D.P. Pappas, M.R. Vissers, D.S. Wisbey, J.S. Kline, J. Gao, Two Level System Loss in Superconducting Microwave Resonators, IEEE Trans. Appl. Supercond., 21 (2011) 871-874.

[27] D.C. Mattis, J. Bardeen, Theory of the anomalous skin effect in normal and superconducting metals, Phys. Rev., 111 (1958) 412-417.

[28] J. Gao, J. Zmuidzinas, A. Vayonakis, P. Day, B. Mazin, H. Leduc, Equivalence of the Effects on the Complex Conductivity of Superconductor due to Temperature Change and External Pair Breaking, J. Low. Temp. Phys., 151 (2008) 557-563.

[29] C.G. Yu, B. Kim, Y.-J. Doh, Fabrication and characterization of magnetic-field-resilient MoRe superconducting coplanar waveguide resonators, Curr. Appl. Phys., 47 (2023) 24-29.

[30] J. Gao, M. Daal, A. Vayonakis, S. Kumar, J. Zmuidzinas, B. Sadoulet, B.A. Mazin, P.K. Day, H.G. Leduc, Experimental evidence for a surface distribution of two-level systems in superconducting lithographed microwave resonators, Appl. Phys. Lett., 92 (2008) 152505.

[31] K. Watanabe, K. Yoshida, T.A. Kohjiro, Kinetic Inductance of Superconducting Coplanar Waveguides, Jpn. J. Appl. Phys., 33 (1994) 5708.

| No. | $s$ ($\mu$m) | $w$ ($\mu$m) | $L_g$ (nH/m) | $C_g$ (pF/m) | $f_c$ (GHz) at 8 mK | $L_k$ (nH/m) | $\alpha_k$ | $Z_0$ ($\Omega$) | $Q_c$ | $Q_i$ | $Q_{TLS,\,0}$ | $Q_{other}$ | $\beta$ | $n_s$ |
|---|---|---|---|---|---|---|---|---|---|---|---|---|---|---|
| D1R1 | 11.5 | 7.0 | 440 | 160 | 3.94835 | 102 | 0.188 | 58.2 | $4.37\times10^3$ | $1.89\times10^3$ | - | - | - | |
| D1R2 | 11.5 | 7.0 | 440 | 160 | 3.98716 | 101 | 0.187 | 58.1 | $5.90\times10^3$ | $5.63\times10^3$ | - | - | - | |
| D1R3 | 11.5 | 7.0 | 440 | 160 | 4.02774 | 109 | 0.199 | 58.6 | $4.37\times10^3$ | $4.95\times10^4$ | $7.61\times10^4$ | $6.65\times10^4$ | 0.82 | 67 |
| D1R4 | 11.6 | 6.7 | 433 | 163 | 4.07825 | 96.6 | 0.182 | 57.0 | $4.83\times10^3$ | $5.92\times10^3$ | $5.14\times10^4$ | $6.24\times10^3$ | 0.82 | 26 |
| D1R5 | 11.6 | 6.9 | 437 | 161 | 4.12380 | 93.5 | 0.176 | 57.4 | $4.74\times10^3$ | $1.22\times10^5$ | - | - | - | |
| D2R1 | 11.9 | 6.5 | 425 | 166 | 4.94704 | 114 | 0.212 | 57.0 | $9.52\times10^3$ | $1.07\times10^5$ | $9.21\times10^4$ | $2.08\times10^5$ | 0.80 | 79 |
| D2R2 | 12.0 | 6.4 | 422 | 167 | 5.37955 | 92.4 | 0.180 | 55.5 | $3.64\times10^3$ | $5.82\times10^4$ | $6.99\times10^4$ | $6.02\times10^3$ | 0.87 | 95 |

**Table 1.** Physical parameters of NbTi SCPW resonators. The geometrical parameters ($s$ and $w$) were obtained from optical microscopy images. The geometrical inductance ($L_g$) and capacitance ($C_g$) were calculated using the conformal mapping method [23]. The resonance frequency ($f_c$), coupled quality factor ($Q_c$), and internal quality factor ($Q_i$) were extracted from transmission spectra fitted to Eq. (1). The kinetic inductance ($L_k$), kinetic inductance fraction ($\alpha_k$), and characteristic impedance ($Z_0$) were calculated from $f_c$. TLS model parameters including the TLS-limited quality factor ($Q_{TLS,0}$), other loss contribution ($Q_{other}$), power dependence exponent ($\beta$), and saturation photon number ($n_s$) were obtained from fits to Eq. (2).

**Figure captions**

**Figure 1.** Pseudo-colored optical microscope image of device D1. (a) Overview of the SCPW chip showing six λ/4 resonators (purple) inductively coupled to a central feedline (green). (b) Enlarged view of a single SCPW resonator. (c) Image of the coupling region between a resonator and the feedline. The light brown regions correspond to the NbTi superconducting film, while the dark brown areas indicate the $SiO_2$ layer on the Si substrate. (d) Magnified view of the central conductor (purple) and ground planes with square-hole arrays.

**Figure 2.** (a) SCPW resonator chip wire-bonded to a PCB board. (b) Aluminum shielding box mounted on the mixing chamber plate. (c) Schematic of the measurement setup for $|S_{21}|$ measurements.

**Figure 3.** $|S_{21}|$ resonance dips measured at $T = 8$ mK for (a) device D1 and (b) device D2. Normalized transmission data $|S'_{21}|$ (black dots) and corresponding fits (red solid line) for (c) resonator D1R3 and (d) resonator D2R1.

**Figure 4.** Insertion loss of (a) D1R3 and (b) D2R1 resonators at $T = 8$ mK for different photon numbers. The photon numbers $n_{ph}$ correspond to input powers $P_{in}$ = -145 dBm (black), -125 dBm (red), -105 dBm (green), and -85 dBm (blue). Photon-number dependence of $Q_i$ (symbols) obtained from (c) D1R3, (d) D1R4, (e) D2R1 and (f) D2R2 resonators at $T = 8$ mK. Solid lines represent fitting results (see text).

**Figure 5.** Temperature dependence of $Q_i$ for (a, c) D1R3 and (b, d) D2R1 resonators. Different colors represent different photon numbers $n_{ph}$. Solid lines denote calculation results (see text).

**Figure 6.** Temperature dependence of the resonance frequency for (a, c) D1R3 and (b, d) D2R1 resonators. Different symbols represent different photon numbers $n_{ph}$. Solid lines represent fitted calculation results (see text).

**Figure 1.**

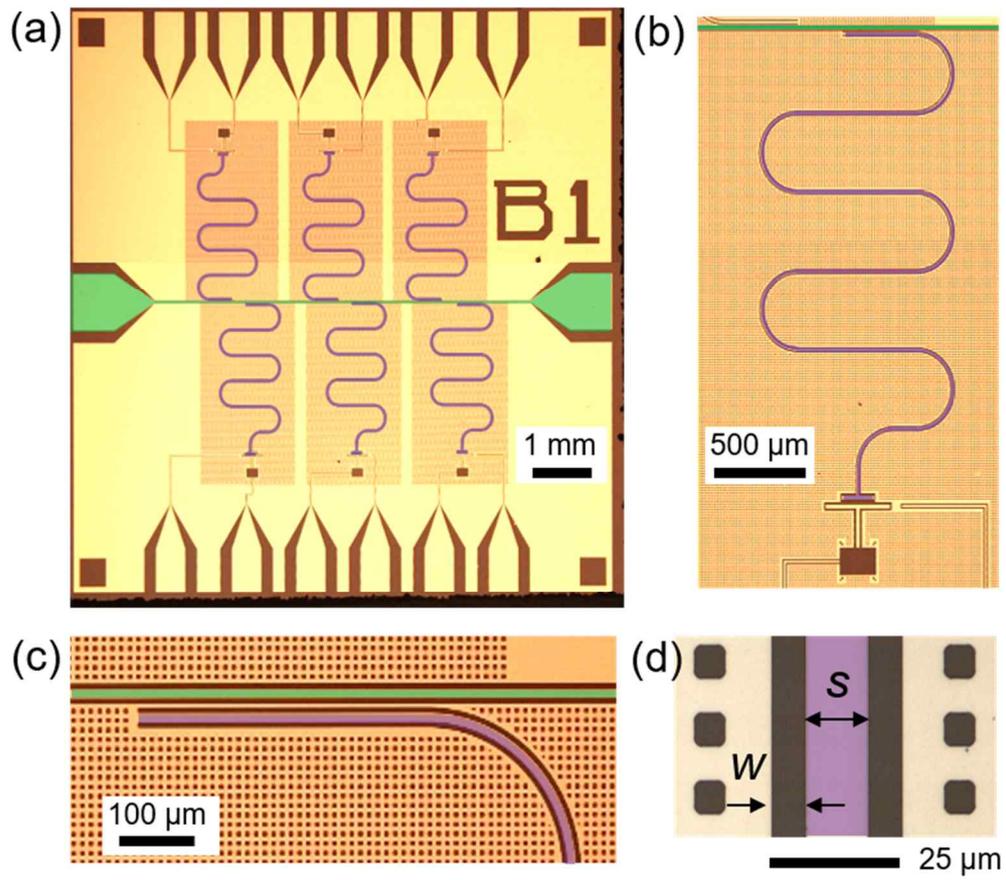

**Figure 2.**

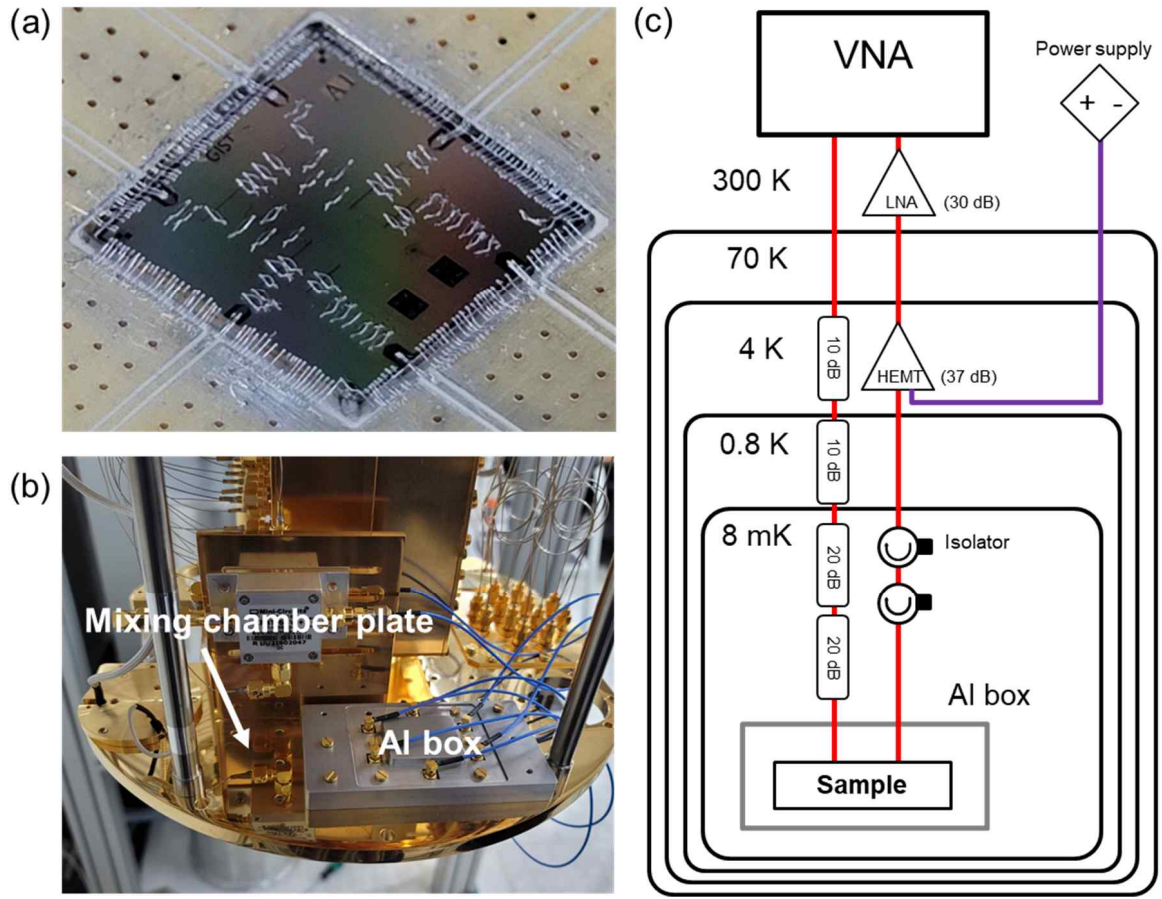

**Figure 3.**

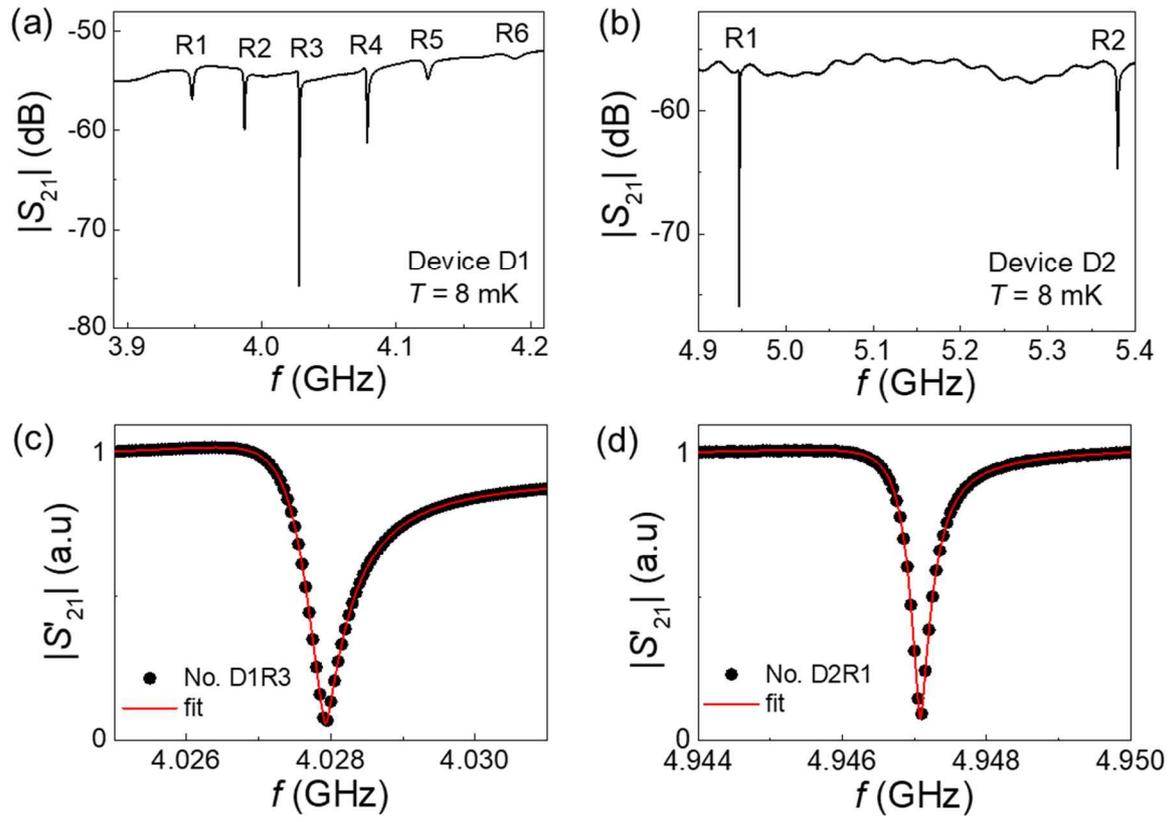

**Figure 4.**

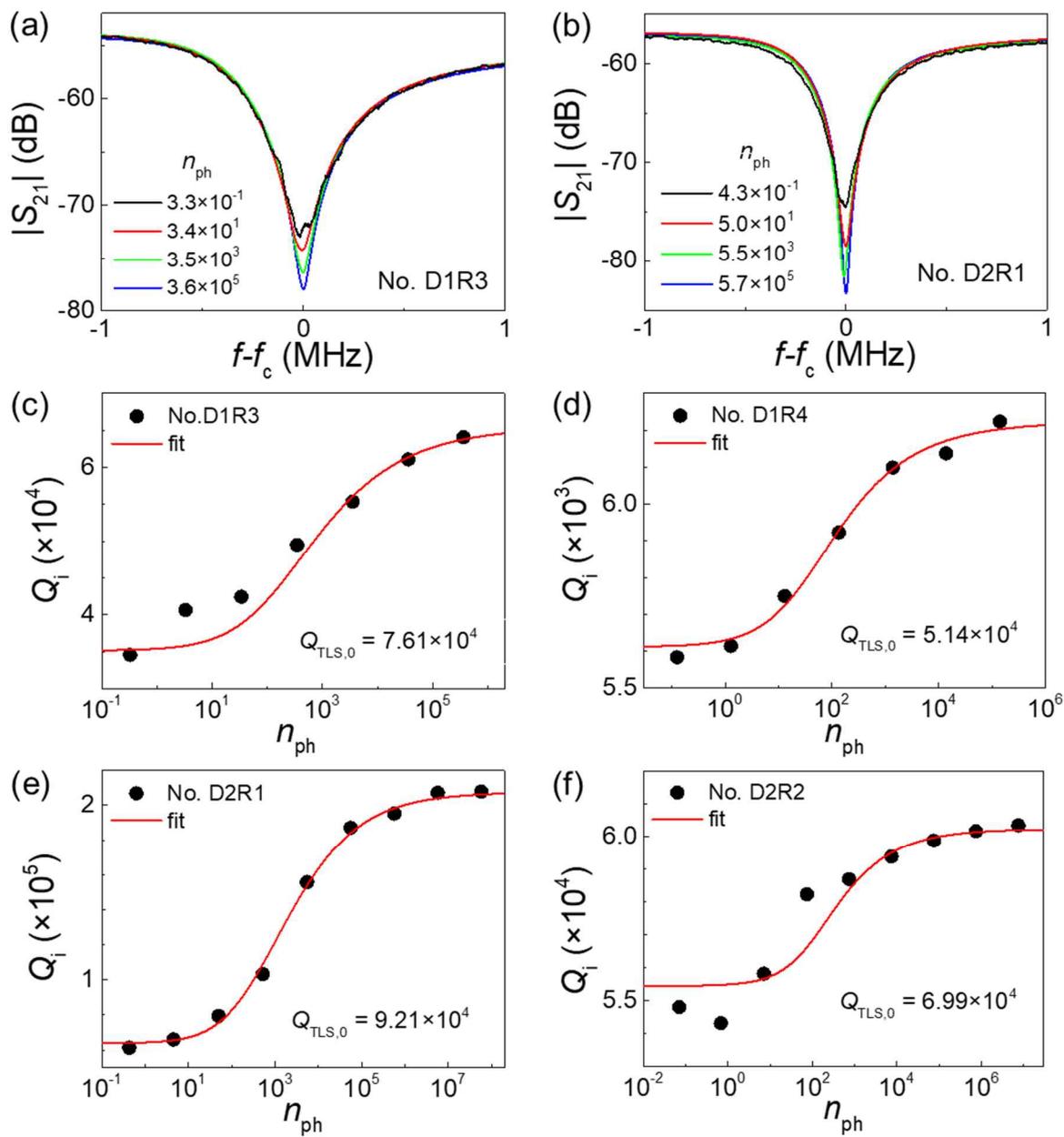

**Figure 5.**

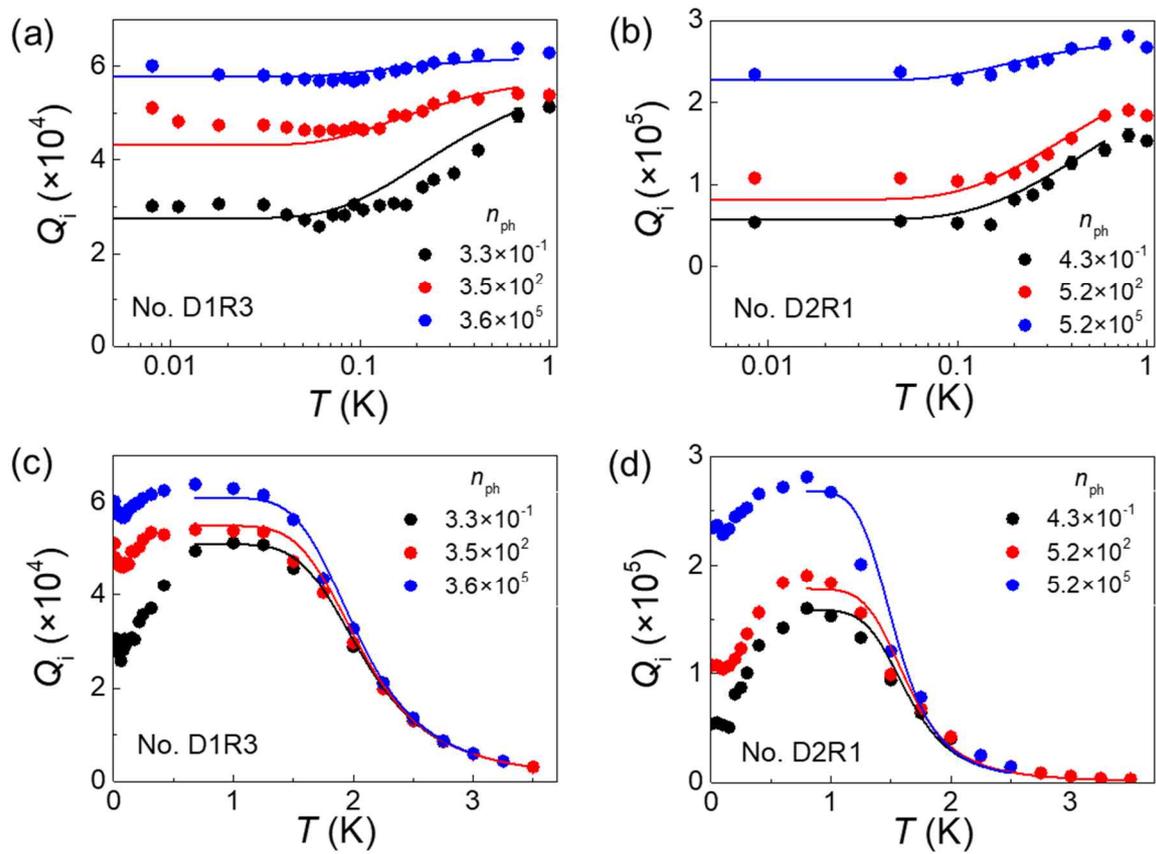

**Figure 6.**

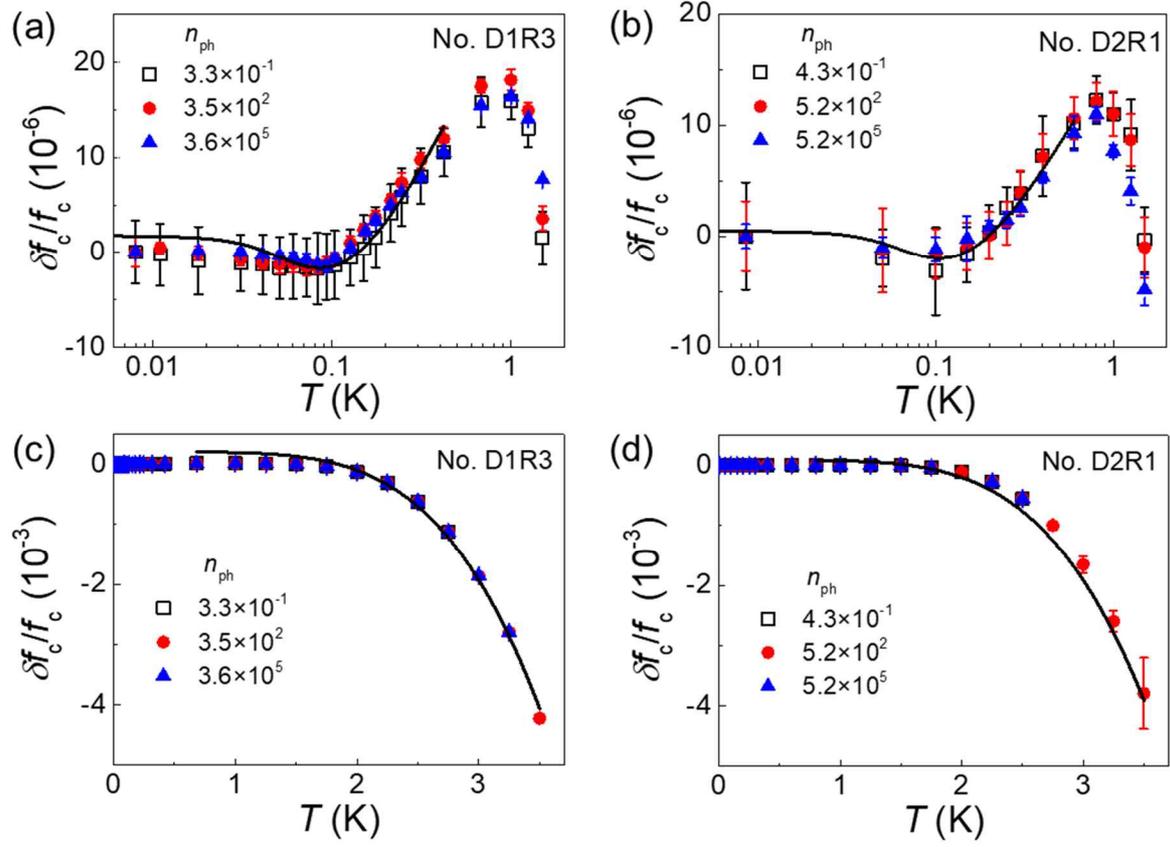